\documentstyle[12pt,epsfig]{article}
\begin{document}
\begin{titlepage}
\begin{center}
{\Large{\bf On the possibilities of distinguishing Dirac from Majorana
neutrinos\vspace{1.5 cm}\footnote{Presented at the XXI Silesian School of Theoretical
Physics on "Recent Progress in Theory and Phenomenology of Fundamental
Interactions", Ustro\'n, September 1997.}}}

{\Large \bf M.~Zra{\l}ek} \\
\vspace{1.5 cm}
{\large{Department of Field Theory and Particle Physics \\
Institute of Physics, University of Silesia \\
Uniwersytecka 4, PL-40-007 Katowice, Poland \\
E-mail: zralek@us.edu.pl}} \\
\vspace{1 cm}
{\bf Abstract} \\
\end{center}
\vspace{0.5 cm}
The problem if existing neutrinos are Dirac or Majorana particles is
considered in a very pedagogical way. After a few historical remarks we
recall the theoretical description of neutral spin 1/2 particles,
emphasizing the difference between chirality and helicity which is 
important in our discussion.
Next we describe the properties of neutrinos in the cases when
their interactions are given by the standard model and by its extensions
(massive neutrinos, right-handed currents, electromagnetic neutrino
interaction, interaction with scalar particles). Various processes where the
different nature of neutrinos could in principle be visible are reviewed. We
clear up misunderstandings which have appeared in last suggestions how to
distinguish both types of neutrinos.
\end{titlepage}

\section{ Introduction.}

The main problem in neutrino physics is the one of the neutrino mass and
mixing between different neutrino flavours. There are many indications that
neutrinos are really massive particles (LSND experiment, problem of the
solar and atmospheric neutrinos, dark matter).

If neutrinos are massive, the next problem is connected with their nature.
Charged fermions are Dirac particles and it is a consequence of the electric
charge conservation. Lepton number conservation is decidedly less
fundamental than charge conservation and it does not govern the dynamics.
Total lepton number can be broken, as it is predicted by many extensions of
the Standard Model (SM). Then neutrinos do not hold any additive internal quantum
numbers and can be identical to their own antiparticles. Such fermions are
now generally known (not only for spin 1/2) as Majorana particles. The
dilemma whether existing neutrinos are Dirac or Majorana particles is the
subject of this paper. We would like to stress that it is not the point if
Dirac and Majorana neutrino differ or not. Of course, they do. Majorana
neutrinos are their own antiparticles, which is not the case for Dirac
neutrino. The problem is whether there is some chance to distinguish them
experimentally (within the Standard Model or beyond the SM neutrino
interactions). These questions can also be divided into two parts. Firstly we
can consider if they are distinguishable in principle and secondly what are
technical possibilities to see different effects in real experiment for both
types of neutrino. We would like to present a critical review of various
efforts and suggestions how to distinguish Dirac from Majorana nature. It is
still a ''hot'' problem and there are many answers emerging here, both
correct and wrong. After short historical remarks (Chapter 2) we remind the
theoretical description of massless and massive, neutral spin 1/2 fermions
(Chapter 3). Next, in Chapter 4, we describe the standard model interaction
of neutrinos and analyze the others, beyond the standard model, neutrino
properties which can give better chance to distinguish their nature. In
Chapter 5 we give a review of various processes where it seems to be
possible for light neutrinos to find some specific signal different for both
characters of neutrinos. In case of wrong suggestions we indicate the
place of errors. In Chapter 6 we summarize our main conclusions.

\section{Historical remarks.}

After Wolfgang Pauli hypothesis [1] in 1930 neutrino was born as a Dirac
fermion described by Paul Dirac equation known from 1927. Neutrino and
antineutrino were distinct particles. Such Dirac particles were used in 1934
by Enrico Fermi in his model of neutrino interactions with nucleons in $%
\beta ^{-}$and $\beta^{+}$ decay [2]. Three years later, in 1937, Etore
Majorana wrote his famous equation [3] in which neutrino was a neutral
object, the same as its own antiparticle. Two years before, Maria Goeppert
Meyer noticed that single $\beta $ decay was not allowed for even-even
nuclei, but decay for such nuclei with emission of two electrons

\begin{equation}
\left( A,Z\right) \longrightarrow \left( A,Z+2\right) +2e^{-}+2\bar{\nu}%
_e, 
\end{equation}
was possible.

Already in 1939 Wendell Furry realized [4] that (if neutrinos have Majorana
character) the neutrinoless double $\beta $ decay

\begin{equation}
\left( A,Z\right) \longrightarrow \left( A,Z+2\right) +2e^{-} 
\end{equation}
would be possible, too.

We will see that this process is also nowadays the best place where the nature
of neutrinos is tested. In 1952 Raymond Davis found no evidence that
antineutrinos from the reactor were absorbed in the chlorine detector by the
reaction [5]

\begin{equation}
\bar{\nu}_e+_{17}^{37}Cl\longrightarrow e^{-}+_{18}^{37}Ar. 
\end{equation}

Four years later in 1956 (after neutrino discovering [6]) it was known that
only neutrino $\nu_e$ could produce electron

\begin{equation}
\nu_e+_{17}^{37}Cl\longrightarrow e^{-}+_{18}^{37}Ar. 
\end{equation}

The results of both observations indicated that neutrino ($\nu_e)$ and
antineutrino $\left( \bar{\nu}_e \right) $ were distinct particles and to
describe the difference the electron lepton number was introduced $\left(
L_{\nu_e}=1,L_{\bar{\nu}_e}=-1\right) .$ After this discovery it was
obvious that the neutrinos should be treated as the Dirac particle 
and not as the Majorana
one.

In 1956 parity violation was discovered by Tsung Dao Lee and Chen-Ning Yang 
[7] and
experimentally supported one year later by Chien-Shiung Wu et al. [8]. 
Immediately it was
realized that breaking of the mirror symmetry is easy to understand if we
assumed that neutrinos were massless particles [9]. Four component spinor,
resolution of the Dirac equation with vanishing mass, decoupled for two
independent two component spinors. The first one, which described particle with
negative and antiparticle with positive helicity $\left(\nu_L,\bar{\nu}%
_R\right),$ and the second one with opposite helicities for particle and
antiparticle $\left( \nu_R,\bar{\nu}_L\right) .$ If the neutrino
interaction was of V-A type then only one particle should be visible, and
experiments should decide which one. Such an experiment had been done in 
1958 by
Maurice Goldhaber et al. in Brookhaven [10]. The answer was clear. The
neutrinos from $\beta ^{+}$ decay had negative helicity and that ones from $%
\beta ^{-}$ were positive helicity states. Only the first $\left( \nu_L,\bar{
\nu}_R\right) $ resolution of massless Dirac equation (known as Weyl equations)
was realized in nature. After this discovery the Davis' result could be
interpreted in an alternative way. The chlorine experiment could only
distinguish negative from positive helicity particle states; it couldnot tell
the difference between Dirac and Majorana neutrinos. So, from the
experimental point of view there was no way to distinguish

\begin{equation}
\begin{array}{c}
\nu_L\Leftrightarrow \nu \left( -\right) , \\  
\\ 
\bar{\nu}_R\Leftrightarrow \bar{\nu}\left( +\right) . 
\end{array}
\end{equation}

In 1957-58 several papers appeared [11] which had shown that there was
equivalence between Weyl and massless Majorana fermions. Then, for almost
twenty years, there was practically no discussion in literature about
neutrino's nature. In the seventies unification theories appeared with
massive neutrinos [12]. The so called ''see-saw'' mechanism made it possible
to understand why the mass of neutrinos was very small [13]. After first
observations of the solar neutrino anomaly [14] the problem of neutrino mass
became one of the most important subjects in particle physics (and later in
astrophysics and cosmology). For massive neutrinos the problem of their
nature once more began to be very important. Fifty years later the Majorana
paper has become again famous, as it poses what Pontecorvo calls ''the central
problem in neutrino physics'': is neutrino identical to its own
antiparticle? From the beginning of eighties papers with different
suggestions how to resolve this problem have been
appearing continuously. Unfortunately, the very pessimistic observation made
in 1982 [15] stating that all observable effects which differentiate Dirac
and Majorana neutrinos disappear if neutrino mass goes to zero is still
valid.

\section{Dirac, Majorana, Weyl neutrinos their helicity, chirality and all
that.}

For the future discussion it is worth presenting a short reminder of
definitions of the basic properties of spin 1/2 fermion.

It is well known that Lorentz group L$_{+}^{\uparrow }$ has two
nonequivalent two-dimensional representations. The objects which transform
under Lorentz transformation are known as the van der Waerden spinors [16],
right $\Psi _{R}$and left $\Psi _L.$

\begin{equation}
\Psi _{R\mbox{ }}
\stackrel{Lorentz\mbox{ }transformation}{\longrightarrow }%
\Psi _{R\mbox{ }}^{^{\prime }}=e^{\frac i2\theta \overrightarrow{n}%
\overrightarrow{\sigma }}e^{-\frac \lambda 2\overrightarrow{m}%
\overrightarrow{\sigma }}\Psi _{R\mbox{ }}, 
\end{equation}
and

\begin{equation}
\Psi _{L\mbox{ }}
\stackrel{Lorentz\mbox{ }transformation}{\longrightarrow }%
\Psi _L^{^{\prime }}=e^{\frac i2\theta \overrightarrow{n}\overrightarrow{%
\sigma }}e^{\frac \lambda 2\overrightarrow{m}\overrightarrow{\sigma }}\Psi
_{L\mbox{ }}, 
\end{equation}
where $\overrightarrow{n},\overrightarrow{m},\theta $ and $\lambda $ are
proper characteristics of Lorentz transformation and $\overrightarrow{\sigma }
$ are Pauli matrices [17]. For zero mass objects these spinors satisfy the
Weyl equations [18].

\begin{equation}
\left( \widehat{\sigma}^{\mu}\partial_{\mu} \right) \Psi _{R\mbox{ }}=0,
\mbox{ }\widehat{\sigma }^\mu =\left( \sigma ^0,
\overrightarrow{\sigma }\right) , 
\end{equation}

\begin{equation}
\left( \sigma ^{\mu}\partial_{\mu}\right) \Psi _L=0,\mbox{ }\sigma ^{\mu}
=\left( \sigma
^0,-\overrightarrow{\sigma }\right) , 
\end{equation}
and describe particle with positive $\left( \Psi _{R\mbox{ }}\right) $ and
negative ($\Psi _L$) helicities. For massless particle the spin projection on
momentum is Lorentz invariant. For particles with mass the Weyl equations
are not satisfied and there are two possibilities. The first one, more
fundamental was discovered by Majorana [3]. The fields $\Psi _{R(L)\mbox{ }}$%
satisfy the Majorana equations

\begin{equation}
i\left( \widehat{\sigma }^{\mu}\partial_{\mu}\right) \Psi _{R\mbox{ }}-m
\varepsilon
\Psi _{R\mbox{ }}^{*}=0, 
\end{equation}
and

\begin{equation}
i\left( \sigma^{\mu}\partial_{\mu}\right) \Psi _{L\mbox{ }}+m^{^{\prime
}}\varepsilon \Psi _{L\mbox{ }}^{*}=0, 
\end{equation}
where m, m$^{^{\prime }}$ are particle masses and $\varepsilon =\left( 
\begin{array}{cc}
0 & 1 \\ 
-1 & 0 
\end{array}
\right).$ Equations (10,11) describe two completely different objects with
masses m and m$^{^{\prime }}$ which do not possess any additive quantum
numbers and particles are their own antiparticles.

The second possibility of the field equation for massive fermion had been known
before Majorana as the Dirac equation [19]

\begin{equation}
\begin{array}{c}
i\left( 
\widehat{\sigma }^{\mu}\partial_{\mu}\right) \Psi _{R\mbox{ }}-m\Psi _{L\mbox{ }}=0,
\\  \\ 
i\left( \sigma^{\mu}\partial_{\mu}\right) \Psi _{L\mbox{ }}-m\Psi _{R\mbox{
}}=0, 
\end{array}
\end{equation}
and had described only one fermion with some additive quantum number (e.g.
charge). Usually this equation is presented in four dimensional Dirac
bispinor formalism as

\begin{equation}
\left( i\gamma ^\mu \partial _\mu -m\right) \Psi =0,
\end{equation}
where
$$
\gamma ^\mu
=\left( 
\begin{array}{cc}
0 & \sigma ^\mu \\ 
\widehat{\sigma }^\mu & 0 
\end{array}
\right) ,\,\,\,\,\,\,\,\,\,\,\mbox{ and \thinspace \thinspace \thinspace
\thinspace \thinspace \thinspace \thinspace }\Psi =
\left( \matrix{\Psi _{R\mbox{ }} \cr 
\Psi _{L\mbox{ }}} \right), 
$$
which is known as Weyl representation for Dirac $\gamma $ matrices. In this
representation let us define

\begin{equation}
\begin{array}{c}
\gamma ^5=i\gamma ^0\gamma ^1\gamma ^2\gamma ^3=\left( 
\begin{array}{cc}
1 & 0 \\ 
0 & -1 
\end{array}
\right) ,\mbox{ P}_L=\frac 12\left( 1-\gamma _5\right) =\left( 
\begin{array}{cc}
0 & 0 \\ 
0 & 1 
\end{array}
\right) , \\  
\\ 
\mbox{and \thinspace \thinspace \thinspace \thinspace \thinspace \thinspace }%
P_R=\frac 12\left( 1-\gamma _5\right) =\left( 
\begin{array}{cc}
0 & 0 \\ 
0 & 1 
\end{array}
\right) . 
\end{array}
\end{equation}
Then%
$$
\Psi _{R\mbox{ }}\equiv \left( \matrix{ \Psi _{R\mbox{ }} \cr 0
} \right) \equiv P_R\Psi ,\mbox{
\thinspace \thinspace \thinspace \thinspace \thinspace \thinspace \thinspace
\thinspace \thinspace }\Psi _{L\mbox{ }}\equiv 
\left( \matrix{ 0 \cr \Psi _{L\mbox{ }%
}} \right) \equiv P_L\Psi . 
$$
The spinors $\Psi _{R(L)\mbox{ }}$ are eigenvectors of $\gamma _5$

\begin{equation}
\gamma _5\Psi _R=\Psi _R,\mbox{ }\gamma _5\Psi _L=-\Psi _L, 
\end{equation}
and are known as chiral eigenvectors with eigenvalues + and -- which have
the name ''chirality''. For massless particles the chirality ''$\pm $'' coincide
with the helicity, $\pm \frac 12.$ For massive particles the chirality and 
helicity
decouple. As we know from Eqs. (6,7) the chirality is Lorentz invariant,
irrespective of whether particle is massive or massless. The helicity is Lorentz
invariant only for massless particle. For a massive particle there always
exist Lorentz frames in which the particle has opposite momentum. This means
that helicity of this particle changes sign and cannot be a Lorentz
invariant object. It is instructive to decompose the free fields $\Psi _{L(R)%
\mbox{ }}$for different kinds of particles in the
helicity representation.

{\scriptsize
\begin{center}
\begin{tabular}{||c||c||c||} \hline \hline
& $m \neq 0$ & m=0 \\ \hline \hline
Dirac & $\Psi_R(x)=\int \left[ A(+)e^{-ikx}-B^{\dagger}(-)e^{ikx} \right]
\chi(+) \sqrt{E+k}$ &
$\Psi_R(x)=\int \left[ A(+)e^{-ikx}-B^{\dagger}(-)e^{ikx} \right]
\chi(+) \sqrt{2E}$ \\
& + $ \int \left[ A(-)e^{-ikx}+B^{\dagger}(+)e^{ikx} \right]
\chi(-) \sqrt{E-k}$ & \\
&& \\
and Weyl & \\
& $\Psi_L(x)=\int \left[ A(-)e^{-ikx}-B^{\dagger}(+)e^{ikx} \right]
\chi(-) \sqrt{E+k}$ &
$\Psi_L(x)=\int \left[ A(-)e^{-ikx}-B^{\dagger}(+)e^{ikx} \right]
\chi(-) \sqrt{2E}$ \\
fields & + $ \int \left[ A(+)e^{-ikx}+B^{\dagger}(-)e^{ikx} \right]
\chi(+) \sqrt{E-k}$ & \\ 
&& \\ \hline
&& \\
Majorana & $\Psi_R(x)=\int \left[ a(+)e^{-ikx}-a^{\dagger}(-)e^{ikx} \right]
\chi(+) \sqrt{E+k}$ &
$\Psi_R(x)=\int \left[ a(+)e^{-ikx}-a^{\dagger}(-)e^{ikx} \right]
\chi(+) \sqrt{2E}$ \\
& + $ \int \left[ a(-)e^{-ikx}+a^{\dagger}(+)e^{ikx} \right]
\chi(-) \sqrt{E-k}$ & \\
&& \\
fields &  $\Psi_L(x)=\int \left[ a(-)e^{-ikx}-a^{\dagger}(+)e^{ikx} \right]
\chi(-) \sqrt{E+k}$ &
$\Psi_L(x)=\int \left[ a(-)e^{-ikx}-a^{\dagger}(+)e^{ikx} \right]
\chi(-) \sqrt{2E}$ \\
& + $ \int \left[ a(+)e^{-ikx}+a^{\dagger}(-)e^{ikx} \right]
\chi(+) \sqrt{E-k}$ & \\ \hline
\end{tabular}
\end{center}}

{\footnotesize Table I. The fields $\Psi_{L(R)}$ for massive Dirac,  
massless Weyl
and for both massive  and massless Majorana neutrinos. See text for all
denotations used in the Table. The integration is over three momentum: 
$\int=\int \frac{d^3k}{(2
\pi)^32E}$.}

\vspace{3 mm}
This decomposition is shown in Table I where we use
the following denotations:

\begin{equation}
\begin{array}{c}
\overrightarrow{k}=k\left( \sin \theta \cos \varphi ,\sin \theta \sin
\varphi ,\cos \theta \right) , \\  \\ 
E=\sqrt{m^2+k^2} 
\end{array}
\end{equation}
are momentum and energy of the particle;

\begin{equation}
\chi \left( \overrightarrow{k},+\right) =\left( 
\begin{array}{cc}
e^{-i\varphi /2} & \cos \theta /2 \\ 
e^{i\varphi /2} & \sin \theta /2 
\end{array}
\right) ,\mbox{ }\chi \left( \overrightarrow{k},-\right) =\left( 
\begin{array}{cc}
-e^{-i\varphi /2} & \sin \theta /2 \\ 
e^{i\varphi /2} & \cos \theta /2 
\end{array}
\right) 
\end{equation}
are Pauli spinors for helicity $+\frac 12$ and $-\frac 12$ respectively, and
the
$A^{\dagger}\left( A\right),$ $B^{\dagger}\left( B\right) $ are creation (annihilation)
operators for Dirac, Weyl particle and antiparticle respectively, and the  $a%
\left( a^{+}\right) $ are suitable operators for Majorana particles.

The careful analysis of the Table I is very instructive and it is worth
making some comments.

1. For the Dirac fields there are two distinct operators, one for particle $%
A^{\dagger}\left( A\right) ,$ and the other one for antiparticle $B^{\dagger}
\left( B\right) $%
. We see that for $E\gg m\neq 0,\;\sqrt{E-k}\approx \frac m{\sqrt{2E}}+{\cal
O}\left(
m^2\right) $ and for definite chirality L or R there are two helicity
states, $h=\pm \frac 12.$ This fact is a consequence of the Lorentz
invariance. However in that case both helicities have
different weights; $\sqrt{E+k}\approx \sqrt{2E}$ for ''good helicity'' and $%
\sqrt{E-k}\approx \frac m{\sqrt{2E}}$ for ''wrong helicity'' states. For
a pure left-handed interaction particles  in the mixed helicity
states will be produced. If helicity is not measured then the chiral particle state with
energy E will be an incoherent superposition of two helicity states
described by the statistical operator $\rho (E)$ 
\begin{equation}
\begin{array}{c}
\rho _{particle}\left( E\right) =\left( 
\frac{E+k}{2E}\right) |\overrightarrow{k}, h =-\frac 12>_p<%
\overrightarrow{k},h =-\frac 12|+ \\ \left( \frac{E-k}{2E}\right) |%
\overrightarrow{k},h =+\frac 12>_p<\overrightarrow{k},h =+\frac
12|. 
\end{array}
\end{equation}
In such a state the neutrino e.g. in $\pi ^{+}\longrightarrow \mu
^{+}\nu{_\mu} $
decay will be produced. It is opposite for antiparticle

\begin{equation}
\begin{array}{c}
\rho _{antiparticle}\left( E\right) =\left( 
\frac{E+k}{2E}\right) |\overrightarrow{k},h=+\frac 12>_a<%
\overrightarrow{k},h =+\frac 12|+ \\ \left( \frac{E-k}{2E}\right) |%
\overrightarrow{k},h=-\frac 12>_a<\overrightarrow{k},h=-\frac
12| 
\end{array}
\end{equation}
as e.g. for the neutrino in $\pi ^{-}\longrightarrow \mu^{-}\bar{\nu}%
_{\mu}$ decay. For relativistic particles the wrong helicity states 
$|\vec{k},h ^{}$ = +$\frac 12>_p$ and $|%
{\vec{k},h }$ = $-\frac 12>_a$ have very small weight $\left(
\frac m{2E}\right) ^2$ and even if, in principle, they can be produced, they
have never been visible. Sometimes they are called ''sterile neutrino''.

2. In the Majorana case there is only one operator which creates particle
and its own antiparticle. Both states (18) and (19) describe the same
object - the Majorana neutrino. There is no sterile neutrino; both helicity
states can be produced with equal weights. The left-handed $\Psi _L$ and
the right-handed $\Psi _R$ fields are connected $\Psi _R(x)=-\varepsilon \Psi
_L^{*}(x).$ The Majorana fields can be also written in the four component
form

\begin{equation}
\Psi (x)=\left( \matrix{ -\varepsilon \Psi _L^{*}(x) \cr \Psi _L(x)} \right), 
\end{equation}
which satisfies the condition

\begin{equation}
i\gamma ^2\Psi ^{*}(x)=\Psi (x),\,\,\,\,\,\,\mbox{\rm where}\mbox{ \thinspace
\thinspace \thinspace \thinspace }i\gamma ^2=\left( 
\begin{array}{cc}
0 & -\varepsilon \\ 
+\varepsilon & 0 
\end{array}
\right) . 
\end{equation}
This relation is sometimes used as a definition of the field for the 
Majorana particle.

3. In cases of both Weyl and massless Majorana neutrinos the limit m$%
\rightarrow $0 is smooth. From Dirac neutrino we obtain two independent Weyl
fields $\Psi _L(x)$ and $\Psi _R(x).$ In the left-handed chiral field $\Psi
_L(x)$ there is particle with negative helicity and antiparticle with
positive helicity. In the field $\Psi _R(x)$ it is just opposite: $A_{%
\overrightarrow{k}}(+)$ and $B_{_{\overrightarrow{k}}}(-)$.

For the massless Majorana neutrino two fields $\Psi _L(x)$ and $\Psi _R(x)$ are
still connected $\left( \Psi _R(x)=-\varepsilon \Psi _L^{*}(x)\right) .$ In
the statical case it was proved [11] that one Weyl neutrino e.g. $\Psi _L(x)$
(or separately $\Psi _R(x)$ ) is equivalent to massless Majorana neutrino
described by two connected fields $\Psi _L$ and $\Psi _R$. This relation is
known as Pauli-Gursey transformation [11] which, for annihilation operators,
can be written in the form

\begin{equation}
\begin{array}{c}
U^{-1}A_{
\overrightarrow{k}}(-)U=a_{\overrightarrow{k}}(-), \\  \\ 
U^{-1}B_{\overrightarrow{k}}(+)U=a_{\overrightarrow{k}}(+), 
\end{array}
\end{equation}
where

\begin{eqnarray}
U&=&exp \left[  \frac{\pi}{4}  \left(  B_{\overrightarrow{k}}^{\dagger}(-)A_{%
\overrightarrow{k}}(-)-A_{\overrightarrow{k}}^{\dagger}(-)B_{\overrightarrow{k}%
}(-)  \right. \right. \nonumber \\
&& \left. \left. -B_{\overrightarrow{k}}^{\dagger}(+)A_{\overrightarrow{k}}(+)+
A_{
\overrightarrow{k}}^{\dagger}(+)B_{\overrightarrow{k}}(+) \right) \right] , 
\end{eqnarray}
and Majorana operators are defined in the following way

\begin{equation}
\begin{array}{c}
a_{
\overrightarrow{k}}(-)=\frac 1{\sqrt{2}}\left[ A_{\overrightarrow{k}}(-)+B_{%
\overrightarrow{k}}(-)\right] , \\  \\ 
a_{\overrightarrow{k}}(+)=\frac 1{\sqrt{2}}\left[ A_{\overrightarrow{k}%
}(+)+B_{\overrightarrow{k}}(+)\right] . 
\end{array}
\end{equation}

It must be stressed that this equivalence theorem is valid only for not
interacting fields. For interacting fields whether the theorem is valid or
not depends on the type of interaction. We will see that the massless
Weyl-Majorana particles are still indistinguishable if there is only left-handed
V-A (or only right-handed V+A) interaction. But for other types of
interactions
this equivalence theorem is no longer true.

\section{Real and hypothetical neutrino interactions.}

\subsection{Neutrinos in the Standard Model.}

It is only one case in which SM predicts masses of particles. The SM
predicts that neutrinos are massless. There are three massless Weyl
neutrinos 
$\nu _e,$ $\nu _\mu $ and $\nu _\tau .$ As a consequence there is no mixing
between generations and (1) leptons have universal interactions, (2) both 
flavour 
$L_e,L_\mu ,$ $L_\tau $ and total $L=L_e+L_\mu +L_\tau $ lepton numbers are
conserved, and (3) there is not CP violation in the lepton sector.

The massless neutrinos have only the left-handed interactions with the 
charged and
neutral gauge bosons

\begin{equation}
L_{CC}=\frac g{2\sqrt{2}}\overline{N}\gamma ^\mu \left( 1-\gamma _5\right)
lW_\mu ^{+}+h.c., 
\end{equation}
and 
\begin{equation}
L_{NC}=\frac g{4\cos \theta_W }\overline{N}\gamma ^\mu \left( 1-\gamma
_5\right) NZ_\mu . 
\end{equation}
The interaction of fermions with the Higgs particles is proportional to the
fermion masses, so massless neutrinos do not interact with scalar particles.

This picture of neutrino interaction is confirmed by all terrestrial
experiments (maybe LSND results are the first which contradict the presented
picture but they still should be better confirmed (e.g. by CARMEN)). In
frame of the SM there is not any chance to differentiate between Weyl and
massless Majorana fermions.

\subsection{The other possible neutrino interactions.}

If neutrinos are massive particles, the mixing between generations appears
in the charged and neutral currents 
\begin{equation}
L_{CC}=\frac g{2\sqrt{2}}\overline{N_a}\gamma ^\mu \left( 1-\gamma _5\right)
K_{al}l_lW_\mu ^{+}+h.c, 
\end{equation}
and 
\begin{equation}
L_{NC}=\frac g{4\cos \theta _W}\overline{N_a}\gamma ^\mu \left( 1-\gamma
_5\right) \Omega _{ab}N_bZ_{\mu}, 
\end{equation}
where $K_{al}\,$and $\Omega _{ab}$ are suitable mixing matrices resulting
from diagonalization of a neutrino mass matrix. If there is a mixing in
the lepton sector then the CP symmetry can be broken. It is the first place which
differentiates the Dirac from Majorana neutrinos.

For the Dirac neutrinos situation looks like in the quark sector. Both charged
leptons and Dirac neutrino fields have the phase transformation freedom

\begin{equation}
N_a\rightarrow N_a^{^{\prime }}=e^{i\alpha _a}N_a \mbox{ \thinspace
\thinspace and  \thinspace \thinspace l}_l\rightarrow \mbox{l}_l^{^{\prime
}}\rightarrow e^{i\beta _l}l_l, 
\end{equation}
and substantial number of phases can be eliminated from mixing matrix K
(matrix $\Omega $ is a function of K).

For the Majorana neutrinos the phase transformation (29) is not 
allowed. We can
see it from the Majorana equation where the field and its complex conjugation are
present simultaneously. Fewer number of phases can be eliminated, so greater amount of
them break the CP symmetry. For example, if neutrinos are Dirac particles we need at
least three families to break CP, for Majorana neutrinos the CP can be
broken already for two families. This different number of CP violating
phases for the Dirac and Majorana neutrino can have real physical consequences
which could be potentially used to distinguish them experimentally. In
practical calculation the difference in CP breaking effects is visible by
the different number of Feynman diagrams. Let assume that the mass of muon
neutrino is larger than the electron neutrino mass $m_{\nu _\mu } > m_{\nu
_e}.$ Then we can calculate the decay width for the process

\begin{equation}
\nu _\mu \rightarrow \nu _e+\gamma . 
\end{equation}
Let us assume that the mixing matrix in Eq. (27) has the form

\begin{equation}
K=\left( 
\begin{array}{cc}
c & se^{i\delta } \\ 
-se^{-i\delta } & c 
\end{array}
\right) . 
\end{equation}
If the neutrinos are Dirac fermions two Feynman diagrams will describe the
process (30) at the one loop level.

\begin{figure}[h]
\epsfig{file=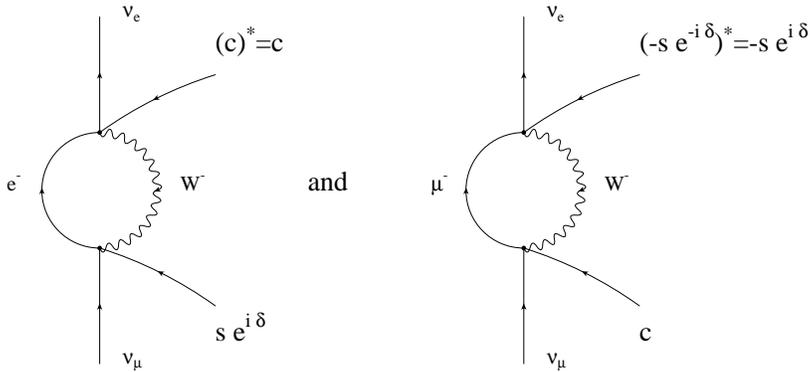,width=13cm}
\caption{
\small{
Two diagrams which describe the radiative neutrino decay 
$\nu_{\mu} \rightarrow \nu_e+ \gamma$ for Dirac neutrinos.}}
\end{figure}

We see from Fig.1 that the CP violating phase $\delta $ multiplies both diagrams
in the same way and cancels after taking modulus square of the sum of both
of them.

In the case of the Majorana neutrino there are four diagrams instead of two
(Fig.2).
\newline
We see that calculating the decay width (Fig.2) the CP violating phase will
not disappear and we obtain [21,22]

\begin{equation}
\Gamma \left( \nu _\mu \rightarrow \nu _e+\gamma \right) \sim \left(
1+\left( \frac{m_e}{m_\mu }\right) ^2+2\left( \frac{m_e}{m_\mu }\right) \cos
2\delta \right) . 
\end{equation}

\begin{figure}[h]
\epsfig{file=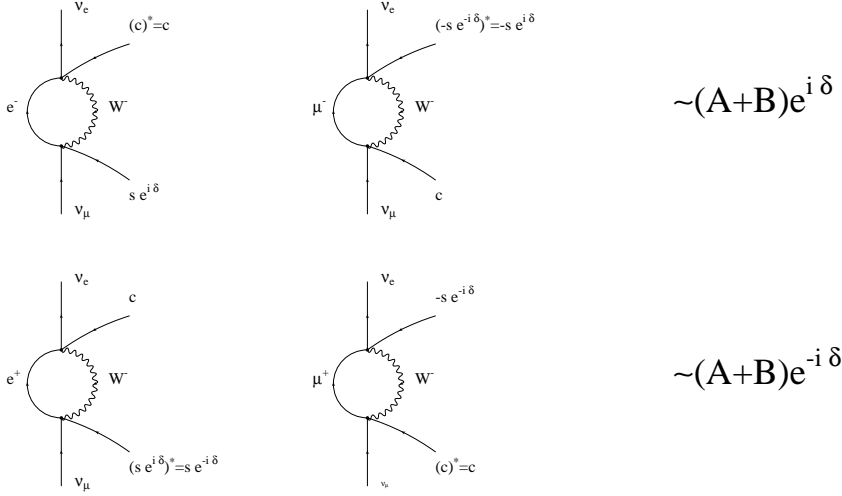,width=13cm}
\caption{\small{Four Feynman diagrams which give contributions to the 
$\nu _\mu
\rightarrow \nu _e+\gamma $ decay for Majorana neutrino.}}
\end{figure}

Even if the CP symmetry is conserved there is an important difference between
the Dirac and Majorana case. For Dirac particles the creation operators for particle 
$(A$) and antiparticle $(B$) can be multiplied by different complex phases,
and , as a result, any CP phase can be absorbed, so it is not physical [23].

For the Majorana neutrino there is only one operator $(A=B$) and the 
CP phase cannot
be absorbed and it is a physical observable. 
The CP phase for the Majorana neutrino must be
pure imaginary number, and for the helicity state there is [23, 24, 25]

\begin{equation}
CP\mbox{ }|\overrightarrow{p},\lambda \rangle \mbox{ }=\eta _{CP}e^{-i\frac
\pi 2}|-\overrightarrow{p},\lambda \rangle , 
\end{equation}
where $\eta _{CP}=\pm i.$

This fact can also have real physical consequences. Let us consider, for
example,
the decay of $\pi^0$  into two identical Majorana
neutrinos $\pi ^0\rightarrow \nu _M\nu _M$ [25]. The initial $\pi ^0$ state is $J^{PC}=0^{-+}$, so the
possible final states are

\begin{equation}
J=0,\mbox{ }S=L=0,\,\,\,\,\,\,\,and\,\,\,\,\,\,\,\,J=0,\mbox{ }S=L=1, 
\end{equation}
so%
$$
CP|\nu _M\nu _M\rangle =\left( \pm i\right) ^2\left( -\right) ^L|\nu _M\nu
_M\rangle \mbox{ = }-|\pi ^0\rangle . 
$$
From this we conclude that $L=0$ so $S=0$ and the emitted neutrino's spins
are antiparallel.

The next important differences between the Dirac and Majorana neutrinos are
their electromagnetic structure. In general any spin 1/2 Dirac fermion can
have four independent electromagnetic formfactors. 

\begin{figure}[h]
\epsfig{file=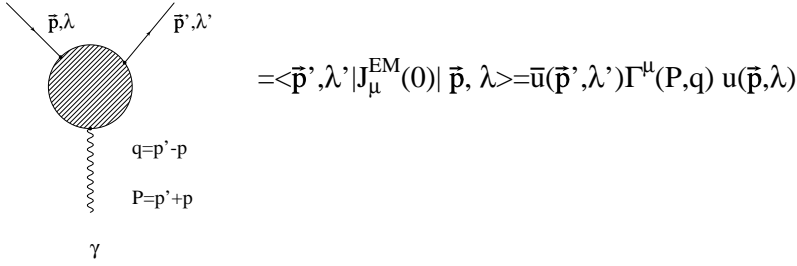,width=18cm}
\caption{\footnotesize{The one photon interaction with a spin 1/2
fermion which is
used to define the electromagnetic structure functions of the Dirac or Majorana
neutrinos.}}
\end{figure}

If we define the one
photon interaction diagram with two fermions like in Fig.3
then the requirements that

(i) initial and final fermions are on shell, 
\newline
and

(ii) the current is conserved $(\Gamma ^\mu q_\mu =0)$ 
\newline
give the structure
function $\Gamma ^\mu \left( P,q\right) $ [26, 27]

\begin{eqnarray}
\Gamma _D^\mu \left( P,q\right)& =&F_D\left( q^2\right) \gamma ^\mu
+iM_D\left( q^2\right) \sigma ^{\mu \nu }q_\nu  \nonumber \\ 
&+& E_D\left( q^2\right) \sigma ^{\mu \nu }q_\nu \gamma _5+G_D\left( q^2\right)
\left( q^\mu 2m-q^2\gamma ^\mu \right) \gamma _5. 
\end{eqnarray}

The structure functions for $q^2\rightarrow 0$ correspond to;

$F_D\left( q^2\right) \stackrel{q^2\rightarrow 0}{\longrightarrow}$
$0,\;\;$
neutrino charge,

$\frac 1{2m}F_D\left( q^2\right) +M_D\left( q^2 \right) 
\stackrel{q^2\rightarrow 0}{\longrightarrow}$ $\mu _m, \;\;$ magnetic moment,

$E_D\left( q^2\right) \stackrel{q^2\rightarrow 0}{\longrightarrow}$ $\mu
_e,\;\;$ electric moment,
\newline
and

$G_D\left( q^2\right) \stackrel{q^2\rightarrow 0}{\longrightarrow}$
$T,\;\;$%
anapol moment [28].

For the Majorana particle only one electromagnetic formfactor survives. There
are several ways to show it:

- the CPT invariance [22, 29],

- identity of fermions in the final state of the decay, $\gamma \rightarrow
\nu _M\nu
_M $ [25, 30], or

- from the Feynman rules - the effective coupling of Majorana fermion with a
neutral vector boson is given by [31] 
\begin{equation}
\Gamma _M^\mu =\Gamma _D^\mu +C\Gamma _D^{\mu _T}C^{-1},
\end{equation}

where

$$
C= \left( \matrix{ -\varepsilon & 0 \cr
           0 & \varepsilon } \right) .
$$

Using any of the above methods
we can show that only the anapol formfactor describes the electromagnetic
structure of the Majorana neutrino

\begin{eqnarray}
\Gamma _M^\mu \left( P,q\right) &=&G_M\left( q^2\right) \gamma ^\mu \gamma
_{5,} \nonumber \\ 
 \mbox{where} \hspace{4 cm} && \nonumber \\ 
G_M\left( q^2\right) &=&-2G_D\left( q^2\right)q^2.
\end{eqnarray}
The electromagnetic structure differentiates the Dirac and Majorana fermions in the
obvious way and we can expect to find some visible experimental effect
connected with this difference.

Besides the diagonal moments (formfactors) which describe the
electromagnetic structure there are also transition moments between
different neutrinos (Fig.4)

\begin{figure}[h]
\epsfig{file=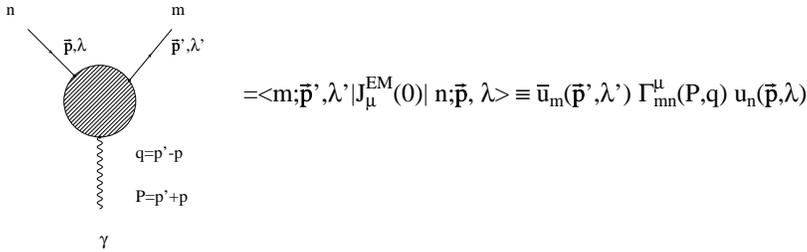,width=18cm}
\caption{\footnotesize{The one photon interaction with two different
fermions which is used to define the transition electromagnetic moments.}}
\end{figure}

Then the CPT symmetry does not give any restrictions for $\Gamma _{mn}^\mu \left(
P,q\right) $ and all four transition moments exist for the Dirac as well as for
the Majorana neutrinos [22,27].
\vspace{5 mm}

Up to now we have discussed the V-A interaction of massive neutrinos.
Experimentally it is not excluded also that a right - handed current
 appears in
the neutrino interaction with the charged and neutral gauge bosons like in
the popular left-right symmetric model [32]. There are also models where
scalar particles interact with neutrinos [33]. Generally it is worth
remembering that 
\begin{equation}
\overline{\Psi }_a\Gamma \Psi _b=\Psi _a^{+}\left( \gamma ^0\Gamma \right)
\Psi _b, 
\end{equation}
from which we can find that for scalar and tensor interactions $\Gamma
=1,\gamma _5,\sigma ^{\mu \nu },\sigma ^{\mu \nu }\gamma _5$ the chirality of $%
\Psi _a$ and $\Psi _b$ must be opposite but for vector interactions $\Gamma
=\gamma ^\mu ,\gamma ^\mu \gamma _5$ the chirality is conserved. It follows very
easily from the properties of the projection operators (Eq. (14))

\begin{equation}
P_LP_R=0,\mbox{ }P_L^2=P_L,\mbox{ }P_R^2=P_R, 
\end{equation}
then e.g.

\begin{equation}
\begin{array}{c}
P_L\gamma ^0\gamma ^\mu P_L\neq 0, 
\mbox{ but }P_R\gamma ^0\gamma ^\mu P_L=0, \\  \\ 
P_L\gamma ^0\gamma _5P_R\neq 0,\mbox{ but }P_L\gamma ^0\gamma _5P_L=0. 
\end{array}
\end{equation}
For relativistic particles, the chirality is almost identical with the
helicity, and we can
transform the above rule for the helicity of incoming and outgoing particles.

\subsection{Differences between the Dirac and Majorana neutrinos for various
neutrino interactions.}

First we consider the situation with vanishing neutrino mass. If there is
only a left-handed interaction $\gamma ^\mu P_L$ (as in the SM), then particles
with negative and antiparticles with positive helicities are produced and
interact after production.

For Weyl particle: $A(-)$ and $B(+)$.

For Majorana object: $a(-)$ and $a(+)$.
\newline
Our interaction will not change $A\longleftrightarrow B$ as well as it will
not change $a(-)\longleftrightarrow a(+)$. We know that the $A$ and the 
$B$ objects interact
differently but we have no possibility to check what is the reason for that.
\newline
That means that two cases

\begin{enumerate}
\item the lepton number is conserved and different helicity in $A$ and $B$
operator has no meaning (so really $A\neq B$ and we have Weyl neutrino), or
\item the interaction depends on the helicity of particles, so $A=B$ and
the particles interact differently because they have different helicities
(Majorana neutrino)
\end{enumerate}
are physically indistinguishable.

In order to answer the question if $A=B$ or not we must have possibility to
compare particle with antiparticle in the same helicity states, so

$$
A(-)\mbox{ with }B(-), 
$$

or

\begin{equation}
A(+)\mbox{ with }B(+). 
\end{equation}
But the operators $A(+)$ and $B(-)$ appear in right-handed chiral state $%
N_R. $ Such a field will appear if there are right-handed currents or scalar
-neutrino interactions. Let us consider a simple example. Beam of massless
neutrinos with negative helicity interacts with matter. We assume that there
is the left-handed and the right-handed current in the neutrinos interaction with
electrons, so

\begin{equation}
L_{CC}=\alpha \overline{l}\gamma ^\mu P_LNW_\mu ^{-}+\beta \overline{l}%
\gamma ^\mu P_RNW_\mu ^{-}+h.c. 
\end{equation}
Then

$\bullet $ for the Weyl neutrino only electrons will be produced in deep
inelastic scattering (Fig.5)

\begin{figure}[h]
\epsfig{file=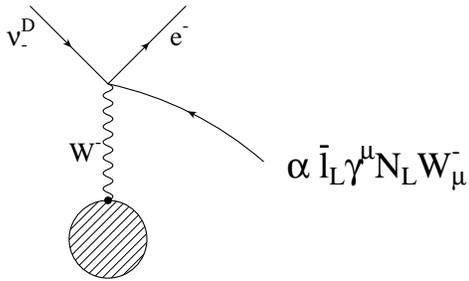,width=8.5cm}
\caption{\footnotesize{The deep inelastic scattering of the massless Weyl
neutrinos with helicity $h=-1/2$. Even if right-handed current exists,
beam of massless neutrinos with negative helicities produces electrons only,
contrary to the case of Majorana neutrinos.}}
\end{figure}

$\bullet $ for the Majorana neutrino electrons and positrons will appear
(Fig.6)

\begin{figure}[h]
\epsfig{file=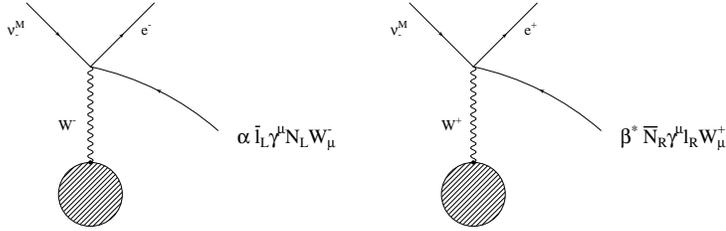,width=11cm}
\caption{\footnotesize{The interaction of the massless negative helicity
Majorana neutrinos with a matter. If right-handed currents exist electrons
and positrons are produced. Electrons (positrons) are produced by the
left- (right-) handed current.}}
\end{figure}

This simple (but unfortunately unrealistic) example shows us that if
the left-handed and the right-handed currents are present we can distinguish
the Weyl
from the massless Majorana neutrinos (if there are the left and right-handed
currents both fields $N_L$ and $N_R$ appear, so we already should talk about
Dirac particle). There is also a more realistic example which convinces us
that the massless Dirac and Majorana neutrinos are distinguishable (at least in
principle) if both left-handed and right-handed currents are present. The
magnetic moment calculated in frame of the L-R symmetric model for Dirac
neutrino does not vanish even if neutrino mass is equal to zero [22, 27, 34]

\begin{equation}
\mu _{\nu_i}=\left( \frac{\sqrt{2}G_F}{\pi ^2}\sin \varphi \cos \varphi
m_e\sum_lm_lRe\left( V_{il}^{+}U_{il}\right) \right) \mu _B 
\end{equation}
(see references for precise denotation of all the parameters in Eq.(43)). 
So,
with vanishing neutrino mass and only left-handed current interaction, there
is no way to distinguish the Dirac from Majorana neutrinos. Such a possibility
appears if neutrinos interact also by right-handed currents or if
interactions with scalar particle are not proportional to the neutrino mass.

The situation will change if neutrinos have some tiny mass. Then even if
there are only left-handed currents, there are ways (at least in principle)
to distinguish both types of neutrino. It is so because the left-handed chiral
states (Eqs. 18,19) are not exactly negative helicity states and, in
principle, there is a possibility to compare interaction of particle and
antiparticle in the same helicity state:

\begin{eqnarray}
\left( \frac m{\sqrt{2E}} \right) |\overrightarrow{k},\lambda & = &
+1/2\rangle _p \;\; \mbox{ for the particle} \nonumber \\
\mbox{\rm with}  \;\;\;\;\;\;\;\;\;\; && \\
 |\overrightarrow{k},\lambda & =& +1/2\rangle _a \;\; \mbox{ for
the antiparticle.} \nonumber
\end{eqnarray}
The several sources of neutrinos are known (reactor, accelerator, the sun,
supernova) but usually they are produced with relativistic energy 
$E \sim 0$ (MeV) 
\{$m_{\nu_e}<3.5\;$%
eV $[35]$ $m_{\nu_{\mu} }<160\;$keV [35] $m_{\nu_{\tau}}<18.2\;$ MeV [36] 
but from
Big-Bang Nucleosynthesis $m_{\nu_{\tau} }<0.95\;$ MeV [37] and from matter density
of the Universe, $m_{\nu_{\tau}} <23\;$ eV [37]\}. Only if we relax the astrophysical
and cosmological information the weight factor $(\frac{m_\tau }{2E_\tau }%
)$ for $\tau \,\,$neutrino is large enough to have interesting value from the experimental
point of view. But, unfortunately, up to now we have not produced the beam
of $\tau $ neutrinos.

Let us consider the interaction of massive Dirac and Majorana neutrinos with
a matter. In both cases the interaction Lagrangian is the same

\begin{equation}
L_{CC}=\frac g{\sqrt{2}}\left\{ \left( \overline{N}\gamma ^\mu P_Ll\right)
W_\mu ^{+}+\left( \overline{l}\gamma ^\mu P_LN\right) W_\mu ^{-}\right\} . 
\end{equation}
If the beam of Dirac neutrinos $\nu ^D$ with helicity $h_\nu $ interacts with a
matter only
electrons with helicities $h_e$ are produced (Fig.7).

\begin{figure}[h]
\epsfig{file=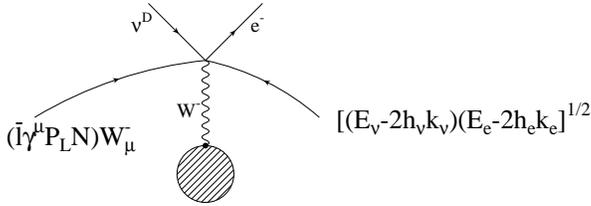,width=11cm}
\caption{\footnotesize{If only left-handed current exists then the beam of
massive Dirac neutrinos will produce electrons. Positrons are not produced
even if neutrinos with positive helicity exist in the beam.}}
\end{figure}

The amplitudes for this process are proportional to the factor  \newline
 $\left[
\left( E_\nu -2h_\nu k_\nu \right) \left( E_e-2h_ek_\nu \right) \right]
^{1/2}$
where $E_\nu \left( E_e\right),$ $k_\nu \left( k_e\right) $ are
energy and momentum of the neutrinos (electrons), respectively. 
Positrons will not be produced
by the Dirac neutrinos even if neutrinos with positive helicity $\left( \nu
_{+}^D\right) $ exist in the beam. Neutrinos in both helicity states will
produce electron only, but mainly $\nu _{-}^D$ will do it, $\nu _{+}^D$
produces e$^{-}\left( h_e=\pm 1/2\right) $ with the small weight $\left(
m_\nu ^2/4E_\nu ^2\right) .$

If the beam of massive Majorana neutrinos interacts with a matter the picture is
different. Both electrons and positrons can be produced (Fig.8)

\begin{figure}[h]
\epsfig{file=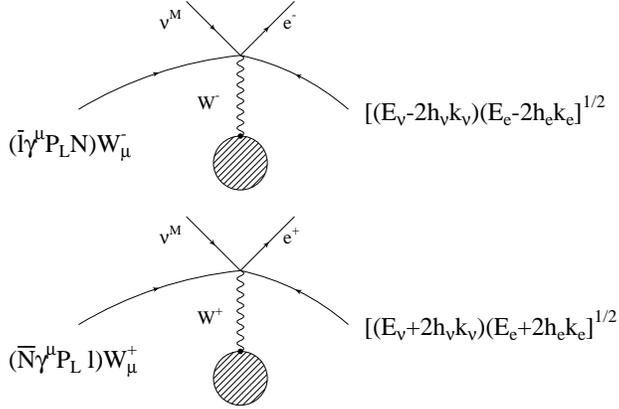,width=11cm}
\caption{\footnotesize{If a beam of massive Majorana neutrinos interacts with
a matter, electrons and positrons can be produced, even if only the
left-handed current describes the neutrino interactions. The second diagram
which is absent for Dirac neutrinos distinguishes both neutrino
characters.}}
\end{figure}

We see that electrons are produced mostly by neutrinos with negative
helicity $\nu _{-}^M,$ contrary to positrons which are produced mainly by $%
\nu _{+}^M.\,$ The Dirac\thinspace and Majorana neutrinos are distinguishable (in
principle) if the second diagram for the Majorana neutrino (absent in the Dirac
case) gives any contribution to the neutrino interaction with a matter. 
It happens
if the factor ($E_\nu +2h_\nu k_\nu )\neq 0,$ and we can conclude that both
neutrinos are distinguishable if $m_\nu \neq 0$ or neutrinos with helicity $%
h_\nu =+1/2$ exist in the beam (even if $m_\nu =0$). The second case means
that the massless Dirac and Majorana neutrinos are distinguishable in the charged
current interaction.

>From presented considerations we could get the impression that massive Dirac
neutrinos are very easily distinguishable from massive Majorana neutrinos.
But it is not the case. In the relativistic beam of neutrinos produced by
the chiral
left-handed interaction the $\nu _{+\mbox{ }}$ neutrinos occur very seldom. 
So,
it is very difficult with the tiny neutrino mass to distinguish Dirac from
Majorana neutrinos using a process where the charged current dominates the
interaction.

What about processes with the neutral currents? There are two things which are
worth considering in this context.

(i) At first sight the neutral current for the Majorana neutrino looks completely
different in comparison with the Dirac neutrino. The Majorana neutrino has no vector
current $\overline{\nu }_M\gamma ^\mu \nu _M=0$ [15, 25].

(ii) In processes where we usually have Dirac neutrino and antineutrino in
Majorana case two identical particles appear. For example in the $Z_0$ decay 
we get two identical Majorana
neutrinos $Z_0\rightarrow \nu _M\nu _M$. Here it seems also that it is easy
to distinguish both cases, because of symetrization procedure for identical
particles.

Let us now consider the first problem. The second one, which was the cause
of many mistakes, we will study in the next Chapter.

From the property (21) for Majorana neutrino it follows that

$$
\overline{\nu }^M\left( x\right) \gamma ^\mu \nu ^M\left( x\right) =0, 
$$
and we have [25, 38]

\begin{equation}
\left\langle \nu _f^M\left| \overline{\nu }^M\gamma ^\mu (1-\gamma _5)\nu
^M\right| \nu _i^M\right\rangle _{\mid x=0}=-\left\langle \nu _f^M\left| 
\overline{\nu }^M\gamma ^\mu \gamma _5\nu ^M\right| \nu _i^M\right\rangle
_{\mid x=0}. 
\end{equation}
If we decompose the neutrino fields in momentum representation (Table I) we
get 
\begin{equation}
\begin{array}{c}
\langle \nu _f^M\left| 
\overline{\nu }^M\gamma ^\mu \left( 1-\gamma _5\right) \nu ^M\right| \nu
_i^M\rangle _{\mid x=0}= \\ \\
\overline{u}\left( \stackrel{\rightarrow }{k}%
_f,h_f\right) \gamma ^\mu \gamma _5\,u\left( \stackrel{\rightarrow }{k%
}_i,h_i\right) +\overline{\upsilon }\left( \stackrel{\rightarrow }{k}%
_i,h_i\right) \gamma ^\mu \gamma _5\upsilon \left( \stackrel{%
\rightarrow }{k}_f,h_f\right) , 
\end{array}
\end{equation}
and from the property 
\begin{equation}
\upsilon =C\overline{u}^T,\mbox{ }\overline{\upsilon }=-u^TC^{-1}, 
\end{equation}
there is 
\begin{equation}
\left\langle \nu _f^M\left| \overline{\nu }^M\gamma ^\mu \left( 1-\gamma
_5\right) \nu ^M\right| \nu _i^M\right\rangle _{\mid _{x=0}}=-2\overline{u}%
_f\gamma ^\mu \gamma _5u_i. 
\end{equation}
For Dirac neutrino at first sight the result is different 
\begin{equation}
\left\langle \nu _f^D\left| \overline{\nu }^D\gamma ^\mu \left( 1-\gamma
_5\right) \nu ^D\right| \nu _i^D\right\rangle _{\mid x=0}=\overline{u}%
_f\gamma ^\mu \left( 1-\gamma _5\right) u_i. 
\end{equation}
But we have to our disposal relativistic neutrinos produced by the left-handed
current so with precision $\left( \frac m{2E}\right) $ the negative helicity
state $\left( h=-1/2\right) $ is the chiral left-handed state (see Eq. 
(18)) so for our initial spinors there is 
\begin{equation}
P_Lu_i\simeq u_i+{\cal O} \left( \frac m{2E}\right) ,\,\,\,\,\,\,\,\,\,P_Ru_i\simeq
{\cal O}\left( \frac m{2E}\right) , 
\end{equation}
from which it follows that 
\begin{equation}
\gamma ^5u_i=-u_i+{\cal O}\left( \frac m{2E}\right) . 
\end{equation}
If we use this relation in (49) and (50) we see that with precision $%
{\cal O}\left( \frac m{2E}\right) $ both matrix elements are equal 
\begin{equation}
\left\langle \nu _f^M\left| \overline{\nu }^M\gamma ^\mu \left( 1-\gamma
_5\right) \nu ^M\right| \nu _i^M\right\rangle =\left\langle \nu _f^D\left| 
\overline{\nu }^D\gamma ^\mu \left( 1-\gamma _5\right) \nu ^D\right| \nu
_i^D\right\rangle +{\cal O}\left( \frac m{2E}\right) . 
\end{equation}
\ \thinspace \thinspace \thinspace \thinspace \thinspace For a tiny neutrino
mass it was also shown that, if there are only the left-handed weak currents,
the electromagnetic structure for the Dirac and Majorana neutrino smoothly
becomes
indistinguishable [15, 39]. As $\sigma ^{\mu \nu }$ and $\sigma ^{\mu \nu
}\gamma _5$ operators in Eq. (35) change chirality and the spinors in
the electromagnetic currents (35) are, with precision $\left( \frac
m{2E}\right) $, the chiral eigenstates $\left( h=-1/2\Longleftrightarrow
P_Ru\cong 0\right) $ the electric and magnetic formfactors must vanish for $%
m_\nu \rightarrow 0$%
\begin{equation}
M_D\left( q^2\right) \stackrel{m_\nu \rightarrow 0}{\longrightarrow }%
0,\,\,\,\,\,\,\,\,\,\,\,\mbox{ }E_D\left( q^2\right) \stackrel{m_\nu
\rightarrow 0}{\longrightarrow }0. 
\end{equation}
For the other formfactors in (35) we have

\begin{eqnarray}
\left\langle \nu _{-}^D\left| J_\mu ^{EM}\right| \nu _{-}^D\right\rangle &=& 
\overline{u}_f(-)\left[ F_D\gamma ^\mu -G_Dq^2\gamma ^\mu \gamma _5\right]
u_i\left( -\right)  \nonumber \\ 
& \approx & \left( F_D+G_Dq^2\right) \overline{u}%
_f(-)\gamma ^\mu u_i\left( -\right) . 
\end{eqnarray}
For the same reasons (only the left-handed current is present) there is no
transition for the $\nu _{+}^D$ Dirac states, so 
\begin{equation}
0\cong \left\langle \nu _{+}^D\left| J_\mu ^{EM}\right| \nu
_{+}^D\right\rangle =\left( F_D-G_Dq^2\right) \overline{u}_f(+)\gamma ^\mu
u_i\left( +\right) , 
\end{equation}
and we have 
\begin{equation}
F_D\approx G_Dq^2. 
\end{equation}
At the same time for the Majorana neutrinos there is (see Eq. (37)) 
\begin{equation}
\left\langle \nu _{-}^M\left| J_\mu ^{EM}\right| \nu _{-}^M\right\rangle
=-2G_Dq^2\overline{u}_f(-)\gamma ^\mu \gamma ^5u_i\left( -\right) =+2G_Dq^2%
\overline{u}_f(-)\gamma ^\mu u_i\left( -\right) . 
\end{equation}
So, if we compare Eqs. (55) and (57) with (58) we see that in the
limit $m_\nu \rightarrow 0$ both electromagnetic currents go smoothly to the
same value 
\begin{equation}
\left\langle \nu _{-}^D\left| J_\mu ^{EM}\right| \nu _{-}^D\right\rangle 
\stackrel{m_\nu \rightarrow 0}{\longrightarrow }\left\langle \nu _{-}^M\left|
J_\mu ^{EM}\right| \nu _{-}^M\right\rangle \stackrel{m_{\nu}\longrightarrow 0}{%
\rightarrow }2G_Dq^2\overline{u}_f\gamma ^\mu u_i. 
\end{equation}
\thinspace In the next Chapter we will see that also the fact that Majorana
particles are indistinguishable from their antiparticles will not help and, 
for the left-handed
interacting neutrino, differences in all observables for the Dirac and Majorana
neutrino smoothly vanish for $m_\nu \rightarrow 0.$ This statement was
formulated in two papers by B. Kayser and R.~Shrock [15] in 1982 and is known
as ''Practical Dirac-Majorana Confusion Theorem''. Since that time many
papers have appeared [40-44] and in the recent time many e-mail texts have
become
available on the hep-ph list [45-50]. They try to find observables where
both neutrinos give the most visible different effects even if their masses
are small [40-42, 44]. Some of them try to find effects in frame of
extensions of the standard model. Some of them are technically correct. Some
of them are not concerned about the practical value of the presented concept
[50]. There are also simply wrong concepts [43, 45, 48].

If there are other neutrino interactions (right-handed currents, interaction
with scalars) observable differences between the Dirac and Majorana neutrinos
could be substantial even for small neutrino mass. But the SM
works very well so effects of any of SM extensions must be small at least for
presently attainable energies.

\section{Review of various processes.}

\subsection{Processes where the differences between Dirac and Majorana
neutrinos are not seen.}

The main neutrino processes which measure their masses do not feel the
differences between the Dirac and Majorana neutrinos

$\diamondsuit $ In all processes which give the bounds on the neutrino mass
e.g. tritium $\beta $ decay $\left( H_1^3\rightarrow H_2^3\rightarrow e^{-}+%
\bar{\nu}_e\right);$ pion and tau decays $\pi ^{+}\rightarrow \mu
^{+}+\nu _\mu $, $\tau ^{-}\rightarrow 2\pi ^{+}3\pi ^{-}(\pi ^0)\nu _\tau $%
, there is only one neutrino which interacts by the charged current. In these
circumstances all differences in measured quantities between the Dirac and
Majorana neutrino disappear.

$\diamondsuit $ In the case of flavour neutrino oscillations, differences
between both types of neutrinos disappear too [51]. It is very easy to see
that. Probability for transition $\nu _\alpha \rightarrow \nu _\beta $ is
given by 
\begin{equation}
P\left( \nu _\alpha \rightarrow \nu _\beta ,t\right) =\left|
\sum_{a=1}^nU_{\beta a}e^{-iE_at}U_{\alpha a}^{*}\right| ^2. 
\end{equation}
The Dirac and Majorana neutrinos give unique signals through the different
structures of the mixing matrices U. There are more CP violating phases for
the Majorana neutrinos. But the formula (60) does not feel these 
additional
phases; they can be eliminated and remaining number of the CP phases is the same
like for the Dirac neutrino.

\subsection{Terrestrial experiments.}

There are many physical observables which feel the difference between Dirac
and Majorana neutrinos. The problem is of course with the size of these
effects.

$\diamondsuit $ In general, any process which violates the total lepton number
(as for example $e^{-}e^{-}\rightarrow W^{-}W^{-}$, $K^{-}\rightarrow \pi
^{+}\mu ^{-}\mu ^{-},$ $K^{-}\rightarrow \pi ^{+}e^{-}e^{-},$ $\nu
\rightarrow \overline{\nu }$ oscillation, ...) will indicate that neutrinos
have the Majorana character. Also the neutrinoless double $\beta $ decay violates
the total lepton number. We will comment on it later.

$\diamondsuit $ There are also processes which do not violate the total lepton
number and occur for Dirac as well as for Majorana neutrinos. But physical
observables (cross sections, angular distributions, energy distributions, 
decay
widths, polarizations) have specific properties which distinguish both types
of neutrino.

For example the angular distributions for the processes like $\nu
e^{-}\rightarrow \nu e^{-},$ $\nu N\rightarrow \nu N,$ $e^{+}e^{-}%
\rightarrow \nu \nu $ look different for the Dirac and Majorana neutrinos [15].

If we could observe, for example, the photon polarization in the process of
neutrino decay $\nu _i\rightarrow \nu _j+\gamma $ then the ratio of the
left-handed $\left( M_L\right) $ to right-handed $\left( M_R\right) $ photon
polarization distinguishes both types of neutrinos [22] 
\begin{equation}
\frac{M_L\left( \nu _i\rightarrow \nu _j+\gamma \right) }{M_R\left( \nu
_i\rightarrow \nu _j+\gamma \right) }= \left\{ 
\matrix{ \left( 
\frac{m_{\nu _j}}{m_{\nu _i}}\right) ^2 \mbox{ for Dirac neutrinos,} \cr
\\ \mbox{
\thinspace \thinspace \thinspace \thinspace \thinspace \thinspace }%
1\,\,\,\,\,\, \mbox{ for Majorana neutrinos.}} \right.
\end{equation}
There are many other processes where the difference can be written (but only
written not observed). Now we consider two examples which were the places
of wrong interpretation in the past [45, 48].

$\star $ Let us exam the scattering process 
\begin{equation}
\nu _\mu +e^{-}\rightarrow \nu _\mu +e^{-}
\end{equation}
which is measured experimentally [52]. 
We assume that a beam of neutrinos is not a pure negative helicity state $\left(
h=-1/2\right) ,$ there is a mixture of $\left( h=+1/2\right) $ and the density
matrix in the helicity basis is 
\begin{equation}
\rho =\left( 
\begin{array}{cc}
\varepsilon & 0 \\ 
0 & 1-\varepsilon 
\end{array}
\right) , 
\end{equation}
where $0 \leq \varepsilon << 1.$

To be general, we take the coupling of neutrinos and electrons with the 
neutral boson $%
Z_0$ in the form 
\begin{equation}
\frac g{2\cos \theta }\left[ \overline{\nu }\gamma ^\mu \left( A_L^\nu
P_L+A_R^\nu P_R\right) \nu +\overline{e}\gamma ^\mu \left(
A_L^eP_L+A_R^eP_R\right) e\right] Z_\mu . 
\end{equation}
Let us define 
\begin{equation}
y=\frac{E_e^{LAB}}{E_\nu ^{LAB}}=\frac 12\left( 1-\cos \theta _{CM}\right) 
\end{equation}
where $E_{e\left( \nu \right) }^{LAB}$ is energy of outgoing electron
(incoming neutrino) in the LAB system, $\theta _{CM}$ is the CM scattering angle.
Then we can calculate the electron LAB energy distribution $\left( m_\nu
\approx 0,\mbox{ }m_e\approx 0\right) $%
\begin{eqnarray}
\frac{d\sigma }{dy}& = &\frac{G_F^2\rho s}{4\pi }\left\{  \left( A_R\right)
^2  \left[ \left( A_R^e\right) ^2+\left( A_L^e\right) ^2\left( 1-y\right)
^2\right] \varepsilon \right. \nonumber \\
&+& \left. \left( A_L\right) ^2  \left[ \left( A_L^e\right)
^2+\left( A_R^e\right) ^2\left( 1-y\right) ^2\right] \left( 1-\varepsilon
\right) \right\}, 
\end{eqnarray}
where%
$$
A_R=\left\{ 
\begin{array}{c}
A_R^\nu \;\;
\mbox{ \thinspace \thinspace \thinspace for Dirac,} \\ A_R^\nu -A_L^\nu
\,\,\,\,\mbox{ for Majorana,} 
\end{array}
\right. 
$$
and 
\begin{equation}
A_L=\left\{ 
\begin{array}{c}
A_L^\nu \,\,\,\,\;\; 
\mbox{ for Dirac,} \\ A_L^\nu -A_R^\nu \mbox{ \thinspace \thinspace
\thinspace \thinspace for Majorana.} 
\end{array}
\right. 
\end{equation}
Such distribution is measured by CHARM collaboration [52] and the result
agrees with the SM \newline
$\left( A_L^l=-1+2\sin ^2\theta _W\mbox{ , }%
A_R^l=-1+2\sin ^2\theta _W\mbox{, }A_L^\nu =1\mbox{, }A_R^\nu =0\mbox{, }%
\varepsilon =0\right) .$ But let us assume that we have better data. Do we
have any chance to distinguish (at least in principle) the Dirac from Majorana
neutrino from the energy distribution (66)? The answer depends on the
polarization of initial neutrino $\left( \varepsilon \right) $ and the
existence of the right-handed currents $\left( A_R^\nu \right) .$

\thinspace $\bullet $ If all initial  $\nu _\mu $ neutrinos are in the
pure state $\left( \varepsilon =0\right) $ and the interaction is pure
left-handed $\left( A_R^\nu =0\right) $ then $\frac{d\sigma ^D}{dy}=\frac{%
d\sigma ^M}{dy}$. This situation we have in the SM.

$\bullet $ If $\varepsilon =0$ but the right-handed current exists,
then there is a difference in normalization of both cross sections

$$\frac{d\sigma ^M}{dy}=\left( 1-\frac{A_R^\nu }{A_L^\nu }\right) ^2\frac{%
d\sigma ^D}{dy},$$
\newline
and the same is true for the total cross section so in principle the 
Dirac and Majorana
neutrinos are distinguishable.

$\bullet $ If $\varepsilon >0$ but there are no the right-handed currents $%
\left( A_R^\nu =0\right) $ both neutrino types are in principle
distinguishable. As $\left( A_L^e\right) ^2\neq \left( A_R^e\right) ^2$ the
energy distribution is different for the Dirac and Majorana neutrinos.

$\bullet $ The best situation is for $\varepsilon >0$ and $A_R^\nu \neq 0$,
there are two factors which change the angular distribution. 

With present
experimental precision there is no chance to measure such details of the energy
distribution. The data agree with the SM $\left( \varepsilon =0\mbox{, }%
A_R^\nu =0\right) $ very well, and only the products of the neutrino and
electron couplings are measured
\begin{equation}
\left( A_L^\nu A_L^e\right) ^2\mbox{ and \thinspace \thinspace }\left(
A_L^\nu A_R^e\right) ^2. 
\end{equation}
Even if we know from other experiments the electron couplings, from neutrino
energy distribution (66) we can find only the sum of the vector 
$g_V^{\nu _{}}$ 
and the axial vector $g_A^\nu \,$couplings $A_L^\nu =g_V^\nu +g_A^\nu $ .We can
say nothing about $g_V^\nu \,$ and g$_A^\nu $ separately [46, 47].
Particularly we cannot conclude that $g_V^\nu \neq 0$ and because of this
muon neutrino is a Dirac particle as it was wrongly suggested in [45].

$\star $ The other interesting problem arises for production of two
neutrinos in the $e^{+}e^{-}$ collision. If the neutrinos are Dirac particles,
neutrino and antineutrino are produced, for Majorana neutrino two
indistinguishable particles appear in the final state. At first sight the
angular distribution of two final neutrinos in the CM frame should look
completely different in both cases. For two identical particles in the final
state $\left( \nu \nu \right) $ the angular distribution must be
forward-backward symmetric, for two different particles $\left( \nu 
\overline{\nu }\right) $ there are no special reasons to have this symmetry.
Two (three) Feynman diagrams give contribution to the process $%
e^{+}e^{-}\rightarrow \nu \overline{\nu }$ $\left( e^{+}e^{-}\rightarrow \nu
\nu \right) $ in the lowest order [53]. Let us specify the momenta and
helicities of the particles 
\begin{equation}
e^{-}\left( p,\sigma \right) +e^{+}\left( \stackrel{}{\overline{p,}}%
\overline{\sigma }\right) \rightarrow \nu \left( k,\lambda \right) +%
\overline{\nu }\left( \overline{k},\overline{\lambda }\right) , 
\end{equation}
and denote $\Delta \sigma =\sigma -\overline{\sigma }$, $\Delta \lambda
=\lambda -\overline{\lambda }.$

\begin{figure}[h]
\epsfig{file=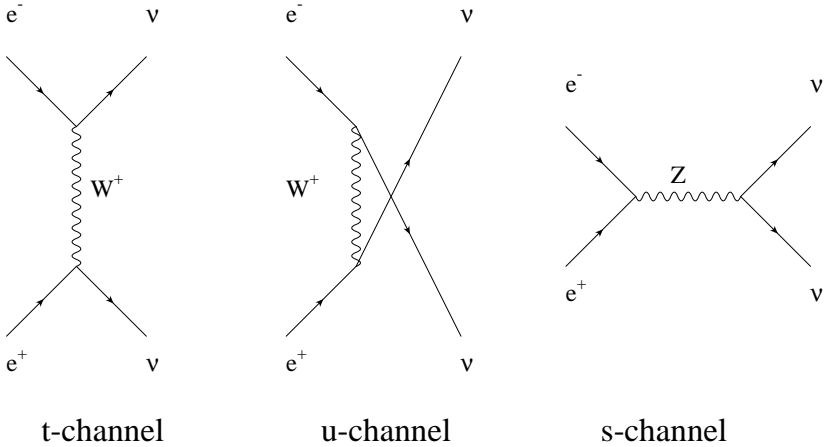,width=11cm}
\caption{\footnotesize{Three Feynman diagrams which contribute to the
$e^{-}e^{+}\rightarrow \nu
\nu $ process for two identical Majorana neutrinos. Only two diagrams, these from t
and s channels, give contribution to the Dirac neutrinos production 
$e^{-}e^{+}\rightarrow \nu \overline{\nu }$ process.}}
\end{figure}

In what follows we use the same denotation for the Z-leptons couplings as in
(Eq.64) but the charged current interaction we take in the form
\begin{equation}
L_{CC}=\frac g{2\sqrt{2}}B_L\overline{\nu }\gamma ^\mu \left( 1-\gamma
_5\right) lW_\mu ^{+}+h.c. 
\end{equation}
If we neglect the lepton masses then only two (four) helicity amplitudes do not
vanish for the Dirac (Majorana) neutrinos [53]. They are (we should remember to
divide the Majorana amplitude by $\sqrt{2}$ for two identical particles in
the final states: $\theta$ is the CM scattering angle) 
\begin{equation}
\begin{array}{c}
M_M\left( \Delta \sigma =1 
\mbox{, }\Delta \lambda =\pm 1\right) =\pm \frac 12f_Re^{i\varphi }\left(
1\pm \cos \theta \right) , \\  \\ 
M_M\left( \Delta \sigma =-1\mbox{, }\Delta \lambda =\pm 1\right) =\pm \frac
12g_L\left( \mp \cos \theta \right) e^{-i\varphi }\left( 1\mp \cos \theta
\right) 
\end{array}
\end{equation}
for the Majorana neutrinos, and 
\begin{equation}
M_D\left( \Delta \sigma =\pm 1\mbox{, }\Delta \lambda =-1\right) =\sqrt{2}%
M_M\left( \Delta \sigma =\pm 1\mbox{, }\Delta \lambda =-1\right) 
\end{equation}
for the Dirac neutrinos ($\Delta \lambda =+1$ neutrinos are sterile in this
case). All other amplitudes are equal to zero.

The $f_R$, $g_L$ parameters are defined by 
\begin{equation}
f_R=\sqrt{2}s\left( \frac e{2\sin \theta _W\cos \theta _W}\right) ^2\frac{%
A_R^eA_L^\nu }{s-M_W^2-i\Gamma _WM_W}, 
\end{equation}
and 
\begin{eqnarray}
g_L\left( \cos \theta \right) & = & \sqrt{2}s  \left\{  
\left( \frac e{\sqrt{2}\sin
\theta _W}\right) ^2\frac{\left| B_L\right| ^2}{\frac s2s\cos \theta -\frac
s2+M_W^2}  \right. \nonumber \\
&+& \left. \left( \frac e{2\sin \theta _W\cos \theta _W}\right) ^2 
\frac{%
A_L^eA_R^\nu }{s-M_W^2-i\Gamma _iM_W}\right\} . 
\end{eqnarray}
The cross section is calculated from the formula 
\begin{equation}
\frac{d\sigma (\Delta \sigma ,\mbox{ }\Delta \lambda )}{d\cos \theta }=\frac
1{32\pi s}\left| M\left( \Delta \sigma ,\mbox{ }\Delta \lambda \right)
\right| ^2, 
\end{equation}
for the polarized particles, and%
$$
\frac{d\sigma }{d\cos \theta }=\frac 1{128\pi s}\sum_{\Delta \sigma ,\Delta
\lambda }\left| M\left( \Delta \sigma ,\mbox{ }\Delta \lambda \right)
\right| ^2, 
$$
in the unpolarized case.

As the t and u amplitudes describe the scattering of Majorana neutrino, the
unpolarized angular distribution is symmetric and completely differs from
the Dirac neutrino distribution which is asymmetric. Does it mean that we have
''very easy'' way of distinguishing the Dirac from Majorana neutrino [48]? Of
course not [49], even if we could measure the angular distribution for the 
outgoing
neutrinos. The point is that all detectors which we have to our disposal are
not able to distinguish the nature of particles from their helicities.
Detection of a particle in direction ($\theta ,\varphi )$ and an antiparticle
in
direction $\left( \pi -\theta ,\mbox{ }\pi +\varphi \right) $ in the CM
frame is technically indistinguishable from the situation that one particle
with helicity $h=-1/2$ travels to the solid angle ($\theta ,\varphi )$ and
the other one to $\left( \pi -\theta ,\mbox{ }\pi +\varphi \right) .$

As an example let us calculate the angular distribution if neutrinos are 
Dirac particles (for simplicity for polarized electron and positron, $\Delta
\sigma =-1).$ Only one amplitude with $\Delta \lambda =-1$ is necessary, and 
\begin{eqnarray}
P_D\left( \Delta \sigma =-1 
\mbox{, }\Delta \lambda =-1;\theta \mbox{,}\varphi \right) & = &
\left| M_D\left(
\Delta \sigma =-1\mbox{, }\Delta \lambda =-1\right) \right| ^2 \nonumber \\
&=&  \left|
g_L(\cos \theta )\right| ^2\frac{\left( 1+\cos \theta \right) }2^2. 
\end{eqnarray}
In the Majorana case, the final particles are identical, and we have to add
incoherently two experimentally indistinguishable situations 
\begin{equation}
\begin{array}{c}
P_M\left( \Delta \sigma =-1 
\mbox{, }\Delta \lambda =-1\right) =\left| M_M\left( \Delta \sigma =-1\mbox{%
, }\Delta \lambda =-1;\theta \mbox{,}\varphi \right) \right| ^2+ \\ \left|
M_M\left( \Delta \sigma =-1 
\mbox{, }\Delta \lambda =+1;\pi -\theta ,\mbox{ }\pi +\varphi \right)
\right| ^2 \\ =2\left| M_M\left( \Delta \sigma =-1\mbox{, }\Delta \lambda
=-1;\theta \mbox{,}\varphi \right) \right| ^2=\left| g_L(\cos \theta
)\right| ^2\frac{\left( 1+\cos \theta \right) }2^2, 
\end{array}
\end{equation}
where we use the relation

\begin{equation}
M_M\left( \Delta \sigma \mbox{, }\Delta \lambda ;\theta \mbox{,}\varphi
\right) =M_M\left( \Delta \sigma, - \Delta \lambda ;\pi -\theta ,%
\mbox{ }\pi +\varphi \right) , 
\end{equation}
which follows from the identity of neutrinos. We see that really two
distributions are identical. The same is true for other electron
polarization (it simply follows from the relations (72) and (78)).
So the total cross section for both electron polarizations are equal%
$$
\sigma _D\left( \Delta \sigma \mbox{, }\Delta \lambda =-1\right) =\sigma
_M\left( \Delta \sigma \mbox{, }\Delta \lambda =-1\right) , 
$$
where the angular distributions for the Dirac (76) and the Majorana 
(77) neutrinos are integrated over full solid angle.

In the Majorana case we can also calculate 
\begin{equation}
P_M\left( \Delta \sigma \mbox{, }\Delta \lambda =+1;\theta \mbox{,}\varphi
\right) =2\left| M_M\left( \Delta \sigma \mbox{, }\Delta \lambda =+1;\theta 
\mbox{,}\varphi \right) \right| ^2 
\end{equation}
which does not exist for Dirac neutrino. But from (72) and (78) it
follows that the distribution (79) is equal 
\begin{equation}
2\left| M_M\left( \Delta \sigma \mbox{, }\Delta \lambda =-1;\pi -\theta ,%
\mbox{ }\pi +\varphi \right) \right| ^2=P_D\left( \Delta \sigma \mbox{, }%
\Delta \lambda =-1;\pi -\theta ,\mbox{ }\pi +\varphi \right) 
\end{equation}
and corresponds to the case of the neutrino which is flying in the direction $%
\pi -\theta ,$ $\pi +\varphi $ and the antineutrino has opposite momentum. So,
for the integrated cross section there is 
\begin{equation}
\sigma _M\left( \Delta \sigma \mbox{, }\Delta \lambda =+1\right) =\sigma
_D\left( \Delta \sigma \mbox{, }\Delta \lambda =-1\right) . 
\end{equation}
For the unpolarized cross section we have to sum (average) over final (initial)
particle polarizations and we get 
\begin{equation}
\frac{d\sigma ^D}{d\cos \theta }=\frac 14\frac 1{32\pi s}\sum_{\Delta \sigma
}P_D\left( \Delta \sigma ,\mbox{ }\Delta \lambda =-1;\theta \right) , 
\end{equation}
and for the Majorana neutrino where also the $\Delta \lambda =+1$ final neutrino
polarization exists 
\begin{equation}
\frac{d\sigma ^M}{d\cos \theta }=\frac{d\sigma ^D}{d\cos \theta }\left(
\theta \right) +\frac{d\sigma ^D}{d\cos \theta }\left( \pi -\theta \right) , 
\end{equation}
which is also obvious as the Majorana case is equivalent to $\nu \left(
h=-1/2\right) +\overline{\nu} \mbox{ }\left( h=+1/2\right) $ for the Dirac
neutrinos. But for the total, unpolarized cross section we once more recover the
equivalence between both types of neutrinos. In order not to take into
account the same spin configuration two times, we have to integrate the
Majorana cross section only over half of the solid angle and we have 
\begin{equation}
\sigma _{tot}\left( M\right) =\int\limits_{-1}^0d\cos \theta \frac{d\sigma ^M%
}{d\cos \theta }=\int\limits_{-1}^{+1}d\cos \theta \frac{d\sigma ^D}{d\cos
\theta }=\sigma _{tot}\left( D\right) . 
\end{equation}
We see that the Practical Dirac-Majorana Confusion Theorem still works.

The main problem in distinguishing Dirac from Majorana neutrino is the lack
of neutrinos with positive helicity. There are two ways, discussed in
literature, how to obtain the neutrino with reversed helicity

(1) to overtake it [54]

(2) to reverse the spin of the neutrino in an external magnetic field [22].

Let us assume that we have the beam of $\pi ^{+}$ with high energy (e.g. 600
GeV from Tevatron) in the laboratory system. In the rest frame of $\pi ^{+}$ the
decay $\pi ^{+}\rightarrow \mu ^{+}+\nu _\mu $ looks like in Fig.10.
All neutrinos with momentum in the backward direction (with respect to the $%
\pi ^{+}$ beam) will have forward momentum in the LAB frame. As a neutrino spin
will not turn after Lorentz transformation, this means that helicity will
change from $h_\nu =-1/2$ in the CM system to $h_\nu =+1/2$ in the LAB system. 

\begin{figure}[h]
\epsfig{file=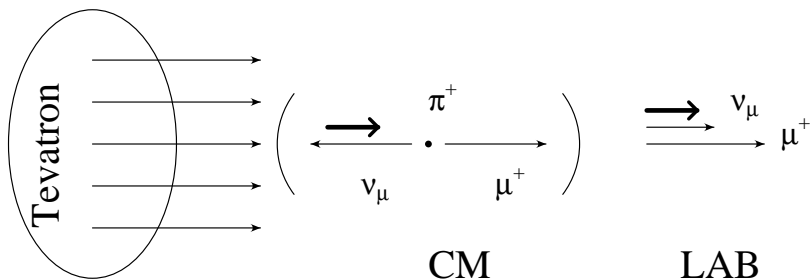,width=11cm}
\caption{\footnotesize{The CM and the LAB frames for decaying pions produced
in the Tevatron. The negative helicity neutrino in the CM frame will change
momentum after Lorentz transformation to the LAB frame so neutrinos with
positive helicities appear in this frame.}}
\end{figure}

This
phenomenon is possible only if neutrino is massive and its efficiency
depends on the neutrino mass. It was shown that m$_{\nu _\mu }\approx \,$10
keV will not be enough if all practical limitations are taken into account
[54]. But there is also some chance that the muon neutrino will oscillate to 
the tau
neutrino which can be much heavier. Now the result depends on the
oscillation probability. It was shown that for U$_{\mu \tau }\approx 0.03$
and m$_{\nu _\tau }>1$ MeV the final results do not look like a completely
wild scheme [55].

The other possibility considered in literature is to reverse the neutrino
helicity by the influence of an external magnetic field [22]. The problem is
that we need large neutrino magnetic moments and large magnetic fields to
obtain visible effect. The SM predicts that for the Dirac neutrino with 
mass $m_\nu $
the magnetic moment is equal [56] 
\begin{equation}
\mu _\nu =\frac{3eG_Fm_\nu }{8\pi ^2\sqrt{2}}\approx 3\times 10^{-19}\left( 
\frac{m_\nu }{1eV}\right) \mu _{Bohr}. 
\end{equation}

Then a magnetic field which is needed for feasible experiment is too large even
for astronomical scale. But present limits on neutrino magnetic
moments are not so small ($\mu _\nu <1.8\times 10^{-10}\mu _B,$ \thinspace
for reactor antineutrino [57]; $\mu _\nu <0.3\times 10^{-11}\mu _B,\,\,\,$%
from stellar cooling [58] ) and stellar magnetic field could have a chance
to reverse neutrino helicity .

There is only one terrestrial experimental approach which currently promises
to state whether neutrinos are Majorana or Dirac particles - it is the
neutrinoless double $\beta $ decay,$\left( \beta \beta \right) _{0\nu }$[4].
The quantity measured in the $\left( \beta \beta \right) _{0\nu }$ is the
average of neutrino masses [59] 
\begin{equation}
\left\langle m_\nu \right\rangle =\sum_nU_{en}^2m_n. 
\end{equation}
If we could find experimentally that $\left\langle m_\nu \right\rangle
>\kappa $ then we would obtain two important items of information, namely
that (i) neutrinos are Majorana type and (ii) at least for one neutrino $%
m_\nu >\kappa .$

Experiments which try to find the amplitudes for the $\left( \beta \beta \right)
_{0\nu }$ decay 
\begin{equation}
\left( Z,A\right) \rightarrow \left( Z+2,A\right) +2e^{-} 
\end{equation}
are presently conducted in several places, using different even-even nuclei.
Up to now the best limit on $\left\langle m_\nu \right\rangle $ has been
obtained by the Heidelberg-Moscow collaboration from observing the half- time of 
$^{76}$Ge [60],%
$$
T_{1/2}^{0\nu }>2.0\times 10^{25}\mbox{ year (68\% of C.L.),} 
$$
which gives%
$$
\left\langle m_\nu \right\rangle <\left( 0.44-1.1\right) \mbox{eV (}68\%%
\mbox{ of C.L.) [61]} 
$$
depending on the method in which the nuclear matrix element is calculated.
\newline
Future experiments which are planed will move the bounds [61]:

Heidelberg-Moscow coll. 
\begin{eqnarray}
^{76}Ge\rightarrow 5\times 10^{25} 
\mbox{ year } & \Rightarrow & \left\langle m_\nu \right\rangle 
\sim 0.2\mbox{ eV}   \\ 
\mbox{NEMO coll.}  \hspace{2.5 cm} &&  \nonumber \\ 
^{100}Mo\rightarrow 10^{25}\mbox{ year } & \Rightarrow & \left\langle m_\nu
\right\rangle \sim 0.16\mbox{ eV.} 
\end{eqnarray}

\subsection{''Half-terrestrial'' experiments.}

Experiments which use neutrinos from non terrestrial sources (the sun, supernova)
were also considered. One such a possibility in which the solar 
neutrinos are used
was recently proposed $\left[ 62\right] $ . If electron neutrino magnetic
moment is in the range of present astrophysical limit $m_\nu <3\times
10^{-12}\mu _B$ there is some chance that strong magnetic fields in the sun
will reverse the neutrino helicity. In [62] the energy distribution 
$\frac{d\sigma}{dT}$ for a final electron in the process 
\begin{equation}
\nu +e^{-}\rightarrow \nu +e^{-} 
\end{equation}
was calculated.As the $\nu _e\left( +\right) $ state is (is not) sterile for
Dirac (Majorana) neutrinos the distribution $\frac{d\sigma }{dT}$ differs
very much for both kinds of neutrinos (the size of the effect depends on the
neutrino density matrix). The electron energy distribution is different for
the Dirac and Majorana neutrino mostly for the low electron energy. So experiments
which can measure the low energy of outgoing electrons are welcome. The
HELLAZ experiment (with the threshold energy 100 keV) seems to be a good
place [62].

It is also possible to find in literature the arguments which compare
neutrino emitted from the supernova with the sun neutrinos [63]. The argument is
as follows. The observation of neutrinos from the SN 1987 A explosion
suggests that the diagonal and the transition moments for Dirac $\nu _e$ neutrino
should be small $\mu _\nu <10^{-13}\mu _B$. Such magnetic moments are
too small to explain hypothetical anticorrelation in the sun neutrino 
observation (visible
in Davis experiment but not in Kamiokande). For explanation of such
anticorrelations we need larger neutrino magnetic moment. So, if neutrinos
are of the Majorana type, both observations could agree.

\section{Conclusions.}

The problem whether the neutrino is identical to its own antiparticle is the
central problem in neutrino physics and very important in particle physics
too.

If the neutrinos have no masses then

$\bullet \,\,$it was proved (Pauli-Gursey transformation) that without
interactions each of Weyl
neutrinos is equivalent with massless Majorana neutrinos,

$\bullet \,$if neutrinos interact with gauge bosons through left-handed
currents (like in the SM) then only one Weyl neutrino appears in the theory
and again we can never distinguish it from the massless Majorana neutrino,

$\bullet \,$if, besides the left-handed also the right-handed currents describe
the gauge bosons --neutrino interaction, or massless neutrinos interact with
scalar bosons then Weyl and Majorana neutrinos interact differently with the
matter. Even if in such case there exist both Weyl neutrinos (equivalent
with one massless
Dirac neutrino), the beam of any kind of Weyl neutrinos behaves in a
different way than the beam of Majorana neutrinos.

If the neutrinos are massive particles, by definition Dirac neutrino ($\nu
\neq \bar{\nu}$) differs from Majorana neutrino ($\nu =
\bar{\nu}$). They interact with the matter in different way
even if only SM governs the neutrino interaction. But unfortunately in such a
case (only left-handed current) all differences in physical observables
disappear in a smooth way with vanishing neutrino mass (Kayser - Shrock
theorem).
\newline
As 
\begin{itemize}
\item \thinspace the SM works very well and any interaction signals
beyond the SM have not been discovered,
\item the neutrino interactions described by the SM are
checked experimentally and are very weak ($\sigma _{\nu e}\approx 10^{-44}$%
\thinspace cm$^2)$, 
\item the masses of light
neutrinos are much smaller than the masses of charged fermions,
\end{itemize}
it is extremely difficult to find clear experimental evidence which informs
us about the nature of existing light neutrinos.

There is only one terrestrial, experimental test that can reveal it, which is
the neutrinoless double beta - decay. Several experimental groups, using
different even - even nuclei, placed the upper limit on the lifetime of the $%
(\beta \beta )_{0\nu }$. Up to now, nobody has found the
evidence for the $(\beta \beta )_{0\nu }$ decay. 
However there are plans to improve
the upper limit for the $(\beta \beta )_{0\nu }$ by almost one order
of magnitude and we can still have hope that the problem if neutrinos are of
the Dirac or the Majorana type can be solved in the nearby future.

\section*{Acknowledgements}

This work was supported by Polish Committee for Scientific Researches under
Grant No. 659/P03/95/08. I thank J. Gluza for reading the text and valuable
remarks.

\newpage\

\end{document}